\documentclass[aps,prb,twocolumn,amsmath,amssymb,%
superscriptaddress,unsortedaddress,%
showpacs]{revtex4}

\usepackage{graphicx}
\usepackage{multirow}
\usepackage{dcolumn}% Align table columns on decimal point
\usepackage{dsfont}

\newcommand{\Avec}[1]{\ensuremath{\mathbf{#1}}}

\begin{document}

\title{A multi-objective optimization procedure to develop
  modified-embedded-atom-method potentials: an application to
  magnesium}

\author{J.~Houze}
\author{Sungho~Kim}
\author{Amitava~Moitra}
\author{B.~Jelinek}
\affiliation{%
  Department of Physics and Astronomy, 
  Mississippi State University,
  Mississippi State, MS 39762, USA
}
\affiliation{%
  Center for Advanced Vehicular Systems, 
  Mississippi State University, 
  Mississippi State, MS 39762, USA
}
\author{Sebastien~Groh}
\author{M.~F.~Horstemeyer}
\affiliation{%
  Center for Advanced Vehicular Systems, 
  Mississippi State University, 
  Mississippi State, MS 39762, USA
}
\affiliation{%
  Department of Mechanical Engineering,
  Mississippi State University,
  Mississippi State, MS 39762, USA
}
\author{Erdem Acar}
\author{Masoud Rais-Rohani}
\affiliation{%
  Department of Aerospace Engineering,
  Mississippi State University,
  Mississippi State, MS 39762, USA
}
\author{Seong-Gon~Kim}\email{kimsg@ccs.msstate.edu} 
\affiliation{%
  Department of Physics and Astronomy, 
  Mississippi State University,
  Mississippi State, MS 39762, USA
}
\affiliation{%
  Center for Advanced Vehicular Systems, 
  Mississippi State University, 
  Mississippi State, MS 39762, USA
}
\affiliation{%
  Center for Computational Sciences, 
  Mississippi State University, 
  Mississippi State, MS 39762, USA
}

\date{\today}

\begin{abstract}

  We have developed a multi-objective optimization (MOO) procedure to
  construct modified-embedded-atom-method (MEAM) potentials with
  minimal manual fitting.  This procedure has been applied
  successfully to develop a new MEAM potential for magnesium.  The MOO
  procedure is designed to optimally reproduce multiple target values
  that consist of important materials properties obtained from
  experiments and first-principles calculations based on
  density-functional theory (DFT).  The optimized target quantities
  include elastic constants, cohesive energies, surface energies,
  vacancy formation energies, and the forces on atoms in a variety of
  structures.  The accuracy of the new potential is assessed by
  computing several material properties of Mg and comparing them with
  those obtained from other potentials previously published.  We found
  that the present MEAM potential yields a significantly better
  overall agreement with DFT calculations and experiments.

\end{abstract}

% PACS numbers

\maketitle

\section{Introduction}

Molecular dynamics simulations are effective tools used to study many
interesting phenomena such as the melting and coalescence of
nanoparticles at the atomic scale.\cite{Shim:2002, Kim:1994:PRL} These
atomistic simulations require accurate interaction potentials to
compute the total energy of the system, and first-principles
calculations can provide the most reliable interatomic potentials.
However, realistic molecular dynamics simulations often require an
impractical number of atoms that either demands too much computer
memory or takes too long to be completed in a reasonable amount of
time.  One alternative is to use empirical or semi-empirical
interaction potentials that can be evaluated efficiently.

The modified-embedded-atom method (MEAM) proposed by Baskes et
al.\cite{Baskes:1989:MEAM, Baskes:1992:MEAM, Baskes:1994:MEAM-hcp} is
one of the most widely used methods using semi-empirical atomic
potentials to date.  The MEAM is an extension of the embedded-atom
method (EAM) to include angular forces.\cite{Daw:1983:EAM,
  Daw:1984:EAM, Baskes:1987:EAM} The MEAM and EAM use a single
formalism to generate semi-empirical potentials that have been
successfully applied to a large variety of materials including fcc,
bcc, hcp, diamond-structured materials and even gaseous elements, to
produce simulations in good agreement with experiments or
first-principles calculations.\cite{Baskes:1987:EAM, Baskes:1989:MEAM,
  Baskes:1992:MEAM, Baskes:1994:MEAM-hcp}

Despite its remarkable successes, one of the most notable difficulties
in using MEAM is that the construction of the MEAM potentials involves
a lot of manual and \textit{ad hoc} fittings.  Because of the complex
relationship between the sixteen MEAM parameters and the resultant
behavior of a MEAM potential, a traditional procedure for constructing
a MEAM potential involves a two-step iterative process.  First, a
\textit{single} crystal structure, designated as the reference
structure, is chosen and the MEAM parameters are fitted to construct a
MEAM potential that reproduces a handful of critical materials
properties of the element in the reference structure.  Second, the new
potential is tested for its accuracy and transferrability by applying
it to atoms under circumstances not used during its construction
phase.  These systems include different crystal structures, surfaces,
stacking faults, and point defects.  If the validation is not
satisfactory, one needs to go back to the first step and adjust the
parameters in a way that improves the overall quality of the
potential.  Although this iterative method does work eventually in
many cases, it is a very tedious and time-consuming.  Ercolessi and
Adams overcame this shortcoming for EAM potentials by developing a
force-matching method that fits the EAM potential to \textit{ab
  initio} atomic forces of many atomic configurations including
surfaces, clusters, liquids and crystals at finite
temperature.\cite{Ercolessi:1994:fm-EAM} Later, the force-matching
method was extended to include many other materials properties such as
cohesive energy, lattice constants, stacking fault energies, and
elastic constants.\cite{Liu:1996:EAM:Mg, Li:2003:PhysRevB.67.125101}
Furthermore, several different MEAM potentials for the same element
often develop and an objective and quantitative method to measure the
relative quality of each potential would be helpful for the
researchers who want to choose one of these potentials.

In this work, we extend the force-matching method to develop a
multi-objective optimization (MOO) procedure to construct MEAM
potentials.  Most realistic optimization problems, particularly in
engineering, require the simultaneous optimization of more than one
objective function.  For example, aircraft design requires
simultaneous optimization of fuel efficiency, payload and weight calls
for a MOO procedure.  In most cases, it is unlikely that the different
objectives would be optimized by the same parameter choices.
Therefore, some trade-off between the objectives is needed to ensure a
satisfactory design.  Stadler\cite{Stadler:1979} introduced the
concept of Pareto optimality\cite{Pareto:1906} to the fields of
engineering and science.  The most widely used method for
multi-objective optimization is the weighted sum method.  A
comprehensive overview and comparison of different MOO methods can be
found in Ref.~\onlinecite{Andersson:2001}.

The composite objective function also provides an unbiased measure to
quantify the relative quality of different MEAM potentials.  We apply
the procedure to develop a new MEAM potential for magnesium.  The new
Mg MEAM potential will be compared with previously published Mg
potentials.

We chose Mg bacause of its increased importantance in many
technological areas, including the aerospace and automotive
industries.  Due to the lower mass densities of magnesium alloys
compared with steel and aluminum and higher temperature capabilities
and improved crash-worthiness than plastics, the use of magnesium die
castings is increasing rapidly in the automotive industry.
\cite{Han:2005:Mg-creep, Lou:1995:Mg-review, Pett:1996:Mg-review}

Empirical potentials for Mg have been previously proposed by several
groups.  In 1988, Oh and Johnson developed analytical EAM potentials
for hcp metals such as Mg.\cite{Oh:1988:EAM} Igarashi, Kanta and
Vitek\cite{Igarashi:1991} (IKV) also developed interatomic potentials
for eight hcp metals including Mg using the Finnis--Sinclair type
many-body potentials.\cite{Finnis:1984} Pasianot and
Savino\cite{Pasianot:1992:PhysRevB.45.12704} proposed improved EAM
potentials for Mg based on IKV's fitting scheme.  Baskes and Johnson
\cite{Baskes:1994:MEAM-hcp} have extended the modified embedded atom
method (MEAM) \cite{Baskes:1987:EAM, Baskes:1989:MEAM,
  Baskes:1992:MEAM} to hcp crystal structures.  Later, Jelinek et al.
improved this potential as a part of the MEAM potentials for Mg-Al
alloy system.\cite{Jelinek:2007:MEAM:Mg-Al} Liu et al. used the
force-matching method to develop an EAM potential for
Mg.\cite{Liu:1996:EAM:Mg}

The paper is organized in the following manner. In
Sec.~\ref{sec:Theory}, we give a brief review of the MEAM. In
Sec.~\ref{sec:Procedure}, the procedure for determination of the MEAM
parameters is presented in detail. In Sec.~\ref{sec:Results}, we
assess the accuracy and transferability of our MEAM potential and make
comparisons to other previously published potentials.

\section{Methodology}
\label{sec:Theory}

\subsection{MEAM}
\label{sec:Theory-MEAM}

The total energy $E$ of a system of atoms in the MEAM
\cite{Kim:2006:MEAM} is approximated as the sum of the atomic energies
\begin{equation}
  E = \sum_{i} E_i.
\end{equation}
The energy of atom $i$ consists of the embedding energy and the pair
potential terms:
\begin{equation}
  E_i = F_i(\bar\rho_{i}) + \frac{1}{2} \sum_{j \neq i}\phi_{ij}(r_{ij}).
\end{equation}
$F_i$ is the embedding function of atom $i$; $\bar\rho_{i}$ is the
background electron density at the site of atom $i$; and
$\phi_{ij}(r_{ij})$ is the pair potential between atoms $i$ and $j$
separated by a distance $r_{ij}$.  The embedding energy
$F_i(\bar\rho_{i})$ represents the energy cost to insert atom $i$ at a
site where the background electron density is $\bar\rho_{i}$. The
embedding energy is given in the form
\begin{equation}
  \label{eq:emb}
  F_i(\bar\rho_{i}) = A_{i} E_{i}^{0} \bar\rho_{i} \ln (\bar\rho_i),
\end{equation}
where the parameters $E_i^0$ and $A_i$ depend on the element type of
atom $i$.  The background electron density $\bar\rho_i$ is given by
\begin{equation}
  \label{eq:rhobar}
  \bar\rho_{i} = \frac{\rho_{i}^{(0)}}{\rho_{i}^0} G(\Gamma_i),
\end{equation}
where
\begin{equation}
  \Gamma_i = \sum_{k=1}^3 \bar{t}_i^{(k)}
  \left(
    \frac{\rho_i^{(k)}}{\rho_i^{(0)}}
  \right)^2
\end{equation}
and
\begin{equation}
  G(\Gamma) = \sqrt{1 + \Gamma}.
\end{equation}
The zeroth and higher order densities, $\rho_i^{(0)}$, $\rho_i^{(1)}$,
$\rho_i^{(2)}$, and $\rho_i^{(3)}$ are given in
Eq.~(\ref{eq:part_den}).  The composition-dependent electron density
scaling $\rho_i^0$ is given by
\begin{equation}
  \rho_i^0 = \rho_{i0}Z_{i0} G( \Gamma_i^\text{ref} ),
\end{equation}
where $\rho_{i0}$ is an element-dependent density scaling, $Z_{i0}$
is the first nearest-neighbor coordination of the reference system, and
$\Gamma_i^\text{ref}$ is given by
\begin{equation}
  \Gamma_i^\text{ref} = \frac{1}{Z_{i0}^2}
  \sum_{k=1}^3 \bar{t}_i^{(k)} s_i^{(k)},
\end{equation}
where $s_i^{(k)}$ is the shape factor that depends on the reference
structure for atom $i$. Shape factors for various structures are
specified in the work of Baskes\cite{Baskes:1992:MEAM}.  The partial
electron densities are given by
\begin{subequations}
  \label{eq:part_den}
\begin{eqnarray}
  \label{eq:part_den_first}
  \rho_i^{(0)} & = & \sum_{j \neq i} \rho_j^{a(0)} S_{ij} \\
  \left( \rho_i^{(1)} \right)^2 & = &
  \sum_{\alpha}
  \left[
    \sum_{j \neq i} \rho_j^{a(1)} \frac{r_{ij\alpha}}{r_{ij}} S_{ij}
  \right]^2 \\
  \left( \rho_i^{(2)} \right)^2 & = &
  \sum_{\alpha, \beta}
  \left[
    \sum_{j \neq i} \rho_j^{a(2)} \frac{r_{ij\alpha}r_{ij\beta}}{r_{ij}^2} S_{ij}
  \right]^2 \nonumber \\
  & & -\frac{1}{3}
  \left[
    \sum_{j \neq i} \rho_j^{a(2)} S_{ij}
  \right]^2
  \\
  \left( \rho_i^{(3)} \right)^2 & = &
  \sum_{\alpha, \beta, \gamma}
  \left[
    \sum_{j \neq i} \rho_j^{a(3)}
    \frac{r_{ij\alpha}r_{ij\beta}r_{ij\gamma}}{r_{ij}^3} S_{ij}
  \right]^2 \nonumber\\
  & & -\frac{3}{5} \sum_{\alpha}
  \left[
    \sum_{j \neq i} \rho_j^{a(3)} \frac{r_{ij\alpha}}{r_{ij}} S_{ij}
  \right]^2,
  \label{eq:part_den_last}
\end{eqnarray}
\end{subequations}
where $r_{ij\alpha}$ is the $\alpha$ component of the displacement vector
from atom $i$ to atom $j$.  $S_{ij}$ is the screening function between
atoms $i$ and $j$ and is defined in Eqs.~(\ref{eq:scr}).  The atomic
electron densities are computed as
\begin{equation}
  \rho_i^{a(k)}(r_{ij}) =
  \rho_{i0} \exp
  \left[
    - \beta_i^{(k)} \left( \frac{r_{ij}}{r_i^0} - 1 \right)
  \right],
\end{equation}
where $r_i^0$ is the nearest-neighbor distance in the single-element
reference structure and $\beta_i^{(k)}$ are element-dependent
parameters.  Finally, the average weighting factors are given by
\begin{equation}
  \bar{t}_i^{(k)} = \frac{1}{\rho_i^{(0)}}
  \sum_{j \neq i} t_{j}^{(k)} \rho_j^{a(0)} S_{ij},
\end{equation}
where $t_{j}^{(k)}$ is an element-dependent parameter.

The pair potential is given by
\begin{align}
  \label{eq:pair}
  \phi_{ij}(r_{ij}) &= \bar\phi_{ij}(r_{ij}) S_{ij} \\
  \begin{split}
  \bar\phi_{ij}(r_{ij}) &= \frac{1}{Z_{ij}}
  \left[ 2E_{ij}^u (r_{ij}) 
    -F_i\left(\frac{Z_{ij}\rho^{(0)}_j(r_{ij})}{Z_i\rho^0_i}\right) \right. \\
    &\quad \left. -F_j\left(\frac{Z_{ij}\rho^{(0)}_i(r_{ij})}
        {Z_j\rho^0_j}\right) \right] \label{eq:rhohat}
\end{split} \\
  E_{ij}^u(r_{ij}) &= -E^0_{ij} \left( 1 + a_{ij}^*(r_{ij})\right)
  e^{-a_{ij}^{*}(r_{ij})} \\
  a_{ij}^{*} &= \alpha_{ij} \left( \frac{r_{ij}}{r_{ij}^0} - 1 \right),
\end{align}
where $\alpha_{ij}$ is an element-dependent parameter.  The
sublimation energy $E^0_{ij}$, the equilibrium nearest-neighbor
distance $r_{ij}^0$, and the number of nearest-neighbors $Z_{ij}$ are
obtained from the reference structure.

The screening function $S_{ij}$ is designed so that $S_{ij} = 1$ if
atoms $i$ and $j$ are unscreened and within the cutoff radius $r_c$,
$S_{ij} = 0$ if they are completely screened or outside the cutoff
radius, and varies smoothly between 0 and 1 for partial screening. The
total screening function is the product of a radial cutoff function
and three-body terms involving all other atoms in the system:
\begin{subequations}
  \label{eq:scr}
  \begin{align}
    \label{eq:scr_first}
    S_{ij} &= \bar{S}_{ij} f_c \left( \frac{r_c - r_{ij}}{\Delta r}
    \right) \\
    \bar{S}_{ij} &= \prod_{k\ne i,j}S_{ikj} \\
    S_{ikj} &= f_c \left(\frac{C_{ikj} - C_{\text{min},ikj}}
      {C_{\text{max},ikj} - C_{\text{min},ikj}} \right) \\
    C_{ikj} &= 1 + 2 \frac{r_{ij}^2 r_{ik}^2 + r_{ij}^2 r_{jk}^2 
      - r_{ij}^4}{r_{ij}^4 - \left( r_{ik}^2 - r_{jk}^2 \right)^2} \\
    f_c\left(x\right) &=
    \begin{cases}
      1 & x \geq 1\\
      \left[ 1 - \left( 1 - x )^4 \right) \right]^2 & 0<x<1 \\
      0 & x \leq 0\\
    \end{cases}
    \label{eq:scr_last}
  \end{align}
\end{subequations}
Note that $C_{\text{min}}$ and $C_{\text{max}}$ can be defined
separately for each $i$-$j$-$k$ triplet, based on their element types.
The parameter $\Delta r$ controls the distance over which the radial
cutoff function changes from 1 to 0 near $r=r_c$.

\subsection{Multi-objective Optimization}
A generic multi-objective optimization (MOO) problem can be formulated
as \cite{Kim:SMO:2005, deWeck:2004}:
\begin{equation}
  \begin{split}
    \min\ &\Avec{J}(\Avec{x}) \quad \text{s.t.} \quad x \in S \\
    \text{where}\ \Avec{J} &=
    [ J_{1}(\Avec{x}) \cdots J_{m}(\Avec{x}) ]^{T} \\
    \Avec{x} &= [ x_{1} \cdots x_{n} ]^{T} \\
  \end{split}
\end{equation}
Here, $\Avec{J}$ is a column vector of $m$ objectives, whereby $J_i
\in \mathds{R}$.  The individual objectives are dependent on a vector
$\Avec{x}$ of $n$ design variables in the feasible domain $S$.  The
design variables are assumed to be continuous and vary independently.
Typically, the feasible design domain is defined by the design
constraints and the bounds on the design variables.  The problem is to
minimize all elements of the objective vector simultaneously.  The
most widely used method for MOO is scalarization using the weighted
sum method.  The method transforms the multiple objectives into an
aggregated scalar objective function $J$ that is the sum of each
objective function $J_i$ multiplied by a positive weighting factor
$w_i$:
\begin{equation}
  J(\Avec{x}) = \sum_{i=1}^{m} w_i J_i(\Avec{x}).
\end{equation}
In this work, the overall goal is to develop a MEAM potential for Mg.
The individual objective functions are constructed from the normalized
differences between the MEAM-generated values and the target values:
\begin{equation}
  \label{eq:Ji(x)}
  J_{i}(\Avec{x}) = \left[ \frac{Q_i(\Avec{x}) - Q_i^{0}}
      {Q_i^{*}} \right]^2.
\end{equation}
Here, $Q_i$ is the physical quantity computed using the current MEAM
potential parameters and $Q_i^{0}$ is the target value to reproduce.
The target values are usually experimental values, but the computed
values from the first-principles method are chosen when the
experimental data are not available.  The normalization factor
$Q_i^{*}$ is a typical value for the given materials parameter and
often $Q_i^{*} = Q_i^{0}$.  The overall objective function
$J(\Avec{x})$ can be minimized using usual multi-dimensional
otimization routines.  To avoid unnecessary complications, we used the
\textit{downhill simplex method},\cite{NumericalRecipes} which
requires only function evaluations, not derivatives.

\section{Potential Construction Procedure}
\label{sec:Procedure}

We used the MOO procedure to develop a new set of MEAM parameters that
improves the overall agreement of MEAM results with experiments or
\textit{ab initio} calculations.  Our previously published MEAM
parameters for Mg\cite{Jelinek:2007:MEAM:Mg-Al} served as the basis
for the present work.

All \textit{ab initio} total-energy calculations and geometry
optimizations are performed within density functional theory (DFT)
using ultrasoft pseudopotentials (USPP) \cite{Vanderbilt:1990} as
implemented by Kresse et. al.\cite{Kresse:1994:VASP:USPP,
  Kresse:1996:VASP:PRB-69} For the treatment of electron exchange and
correlation, we use local-density approximation
(LDA)\cite{Ceperley:1980, Perdew:1981}.  The Kohn-Sham equations are
solved using a preconditioned band-by-band conjugate-gradient (CG)
minimization.\cite{Kresse:1993:PhysRevB.47.558} The plane-wave cutoff
energy is set to at least 300~eV in all calculations.  Geometry
relaxations are performed until the energy difference between two
successive ionic optimizations is less than 0.001~eV.  The Brillouin
zone is sampled using the Monkhorst-Pack
scheme\cite{Monkhors:1976:PhysRevB.13.5188} and a Fermi-level smearing
of 0.2~eV was applied using the Methfessel-Paxton
method.\cite{Methfessel:PhysRevB.40.3616}

The objectives used in this work include equilibrium hcp lattice
constants $a$ and $c$ at 0~K, the cohesive energy, elastic constants,
vacancy formation energy, surface energies, stacking fault energies,
and adsorption energies.  We also used the forces on Mg atoms in
structures equilibriated at six different temperatures.  The final
MEAM parameters obtained from the MOO procedure are listed in
Table~\ref{tab:MEAM-Mg}.  Table~\ref{tab:objectives} shows the
complete list of objectives optimized to construct the MEAM potential
parameters for Mg and their weights.

\subsubsection{Cohesive energies}

The cohesive energy of Mg atom is defined as the heat of formation per
atom when Mg atoms are assembled into a crystal structure:
\begin{equation}
  E_\text{coh} = \frac{E_{\text{tot}} - N E_{\text{atom}}}{N},
  \label{eq:E_c}
\end{equation}
where $E_{\text{tot}}$ is the total energy of the system, $N$ is the
number of Mg atoms in the system, and $E_{\text{atom}}$ is the total
energy of an isolated Mg atom.  The cohesive energies of Mg atoms in
hcp, fcc, and bcc crystal structures for several atomic volumes near
the equilibrium atomic volumes were calculated.  Fig.~\ref{fig:E-V-Mg}
is an example of the cohesive energy plot of Mg atoms as a function of
the lattice constant.  The minimum of this curve determines the
equilibrium lattice constant $a_0$ and equilibrium cohesive energies
$E_\text{hcp}= E_\text{hcp}$ in Table~\ref{tab:objectives}.

\subsubsection{Elastic constants}

Hexagonal crystals have five independent elastic constants: $C_{11}$,
$C_{12}$, $C_{13}$, $C_{33}$, and $C_{44}$.\cite{Ledbetter:1977} The
elastic constants are calculated numerically by applying small strains
to the lattice.  For small deformations, the relationship between
deformation strain and elastic energy increase in an hcp crystal is
quadratic:\cite{Liu:1996:EAM:Mg}
\begin{enumerate}
\item $\Delta U = \delta^2(C_{11} + C_{12})$, for deformation $x' =
  x+\delta\cdot x$, $y'=y+\delta\cdot y$,
\item $\Delta U = \delta^2(C_{11} - C_{12})$, for deformation $x' =
  x+\delta\cdot x$, $y'=y-\delta\cdot y$,
\item $\Delta U = \delta^2 C_{33}/2$, for deformation $z' =
  z+\delta\cdot z$,
\item $\Delta U = \delta^2 (2C_{11}+C_{33}+2C_{12}+4C_{13})/2$, for
  deformation $x' = x+\delta\cdot x$, $y'=y+\delta\cdot y$, $z' =
  z+\delta\cdot z$,
\item $\Delta U = \delta^2 C_{44}/2$, for deformation $z' =
  z+\delta\cdot x$,
\end{enumerate}
where unprimed (primed) are the coordinates of the lattice before
(after) deformation.  $\Delta U$ is the elastic energy due to the
deformation, and $\delta$ is the small strain applied to the lattice.
We follow the procedure described by Mehl et al.\cite{Mehl:1994} and
apply several different strains ranging from $-2.0$\%\ to $+2.0$\%.
The elastic constants are obtained by fitting the resultant curves to
quadratic functions.  We found that this method gives much more stable
results than using one strain value\cite{Liu:1996:EAM:Mg}.

\subsubsection{Surface formation energies}

Surface formation energy per unit surface area is defined as
\begin{equation}
  \gamma = \left(E_{\text{tot}} - N\varepsilon\right)/A,
\end{equation}
where $E_{\text{tot}}$ is the total energy of the system with a
surface, $N$ is the number of atoms in the system, $\varepsilon$ is
the total energy per atom in the bulk, and $A$ is the surface area.
Table~\ref{tab:objectives} lists the surface formation energies used
in this study.  The (10$\bar{1}$0) surface of hcp crystals can be
terminated in two ways, either with a short first interlayer distance
$d_{12}$ (``short termination'') or with a long $d_{12}$ (``long
termination'') (See, for example, Fig.~2 of
Ref.~\onlinecite{PhysRevB.53.13715}).  In this study, we only included
the results for the short terminated surface, since it is known to be
enegetically more favorable over the long terminated surface
\cite{PhysRevB.60.15613} in agreement with experimental observations
in Be(10$\bar{1}$0) and other hcp metals.\cite{PhysRevB.53.13715}

\subsubsection{Stacking fault energies}

Stacking fault formation energy per unit area is defined by
\begin{equation}
  \label{eq:E_sf}
  E_{\text{sf}} = \left(E_{\text{tot}} - N\varepsilon\right)/A,
\end{equation}
where $E_{\text{tot}}$ is the total energy of the structure with a
stacking fault, $N$ is the number of atoms in the system,
$\varepsilon$ is the total energy per atom in the bulk, and $A$ is the
unit cell area that is perpendicular to the stacking fault.  For Mg,
four stacking fault types from the calculation of Chetty et
al.\cite{Chetty:1997:sf-in-Mg} were examined.  The sequences of the
atomic layers within the unit cell of our simulations are: $I_1=ABABAB
CBCBCB$, $I_2=ABABAB CACACB$, $T_2=ABABAB CBABAB$, and $E=ABABAB
CABABAB$.  We note that the unit cells for $I_1$ and $I_2$ contain
\textit{two} stacking faults and the quantities obtained from
Eq.~(\ref{eq:E_sf}) must be divided by two to obtain the correct
formation energies.

\subsubsection{Vacancy formation energies}

The formation energy of a single vacancy $E_{\text{vac}}$ is defined
as the energy cost to create a vacancy:
\begin{equation}
  E_{\text{vac}} = E_{\text{tot}}[N] - N\varepsilon,
  \label{eq:E^f_v}
\end{equation}
where $E_\text{tot}[N]$ is the total energy of a system with $N$ atoms
containing a vacancy, and $\varepsilon$ is the energy per atom in the
bulk.

\subsubsection{Atomic Forces}

For forces, the objective functions are defined as:
\begin{equation}
  J_{i}(\Avec{x}) = \frac{(\left<(\Avec{F}-\Avec{F}^{0})^2\right>)^{1/2}}
  {(\left<(\Avec{F}^{0})^2\right>)^{1/2}},
\end{equation}
where $\Avec{F}$ are the force vectors on atoms calculated using the
MEAM while $\Avec{F}^{0}$ are the force vectors from DFT method.
$(\left<(\Avec{F}^{0})^2\right>)^{1/2}$ represents the
root-mean-square of the DFT force, and
$(\left<(\Avec{F}-\Avec{F}^{0})^2\right>)^{1/2}$ is the
root-mean-square of the error in the force.

To obtain the force data, initial atomic structures that contain 180
Mg atoms were created from the bulk hcp crystal structure.  The
positions of atoms are randomly disturbed from their equilibrium
positions and $10\,000$ steps of molecular-dynamics (MD) simulations
with a timestep of $\Delta t$ = 2.5~ps were performed to equilibrate
each structure for different temperatures.  In each MD run, we used Mg
MEAM potential by Jelinek et al.\cite{Jelinek:2007:MEAM:Mg-Al} If no
MEAM potential were available for MD simulations, one could use an
intermediate MEAM potential that is generated with this MOO procedure
without the force data.  The potential should be adequate enough to
obtain a reasonable set of structures.

\section{Results and Discussion}
\label{sec:Results}

The hcp structure was chosen as the reference structure for Mg.  The
final MEAM parameters obtained from the MOO procedure are listed in
Table~\ref{tab:MEAM-Mg}.
\begin{table*}[!tbp]
  \caption{\label{tab:MEAM-Mg} Set of the MEAM potential parameters 
    for Mg. See Sec.~\ref{sec:Theory-MEAM} for the definition of
    these parameters and their usage. The hcp structure was chosen 
    as the reference structure for Mg.}
  \begin{ruledtabular}
    \begin{tabular}{cccccccccccccccc}
      $E^{0}$[eV] & $a_0$[\AA] & $A$ & $\alpha$ &
      $\beta^{(0)}$ & $\beta^{(1)}$ & $\beta^{(2)}$ & $\beta^{(3)}$ &
      $t^{(0)}$ & $t^{(1)}$ & $t^{(2)}$ & $t^{(3)}$ &
      $C_{\text{max}}$ & $C_{\text{min}}$ &
      $r_c$ & $\Delta r$ \\
      \hline
      1.51  & 3.20 &  1.14 & 5.69 &
      2.66  & -0.003  &  0.348 & 3.32  &
      1.00  & 8.07  & 4.16 & -2.02  &
      3.22   & 1.10 & 5.0 & 0.353 \\
    \end{tabular}
  \end{ruledtabular}
\end{table*}

\subsection{Materials properties}

\begin{table*}[!tbp]
  \caption{\label{tab:objectives} The objectives optimized to
    construct the MEAM potential parameters for Mg along with
    the target values and their weights.  Comparisons
    to other Mg potentials are also made.  $E_{\text{coh}}$ is the
    cohesive energy, $B$ is the bulk modulus, $\gamma$ is the
    surface energy, 
    $E_{\text{sf}}$ is the stacking fault formation energy, 
    $E_{\text{vac}}$ is the relaxed vacancy formation energy, and
    $\Delta \Avec{F}$ is the relative error in forces.  
    The underlined quantities are the target values chosen for the 
    MOO procedure.  
  }
  \begin{ruledtabular}
    \begin{tabular}{clccccccccc}
      & Objective & Unit & Weight & Expt & DFT
      & MEAM\footnotemark[1] & Jelinek\footnotemark[2]
      & Liu\footnotemark[3] & Hu\footnotemark[4] \\
      \hline
      1 & $a_0$ & \AA & 1.0 
      & \underline{3.21} [\onlinecite{Emsley:1998}] & 3.128
      & 3.21 & 3.21 & 3.21 \\
      2 & $c/a$ & - & 1.0 
      & \underline{1.623} [\onlinecite{Emsley:1998}] & 1.623 
      & 1.622 & 1.623  & 1.623  & \\
      3 & $E_\text{coh} = E_\text{hcp}$ & eV & 2.0 
      & \underline{1.51} [\onlinecite{Kittel:1996}] & 1.78
      & 1.51  & 1.55  & 1.52  & \\
      4 & $B$ & kbar & 1.0 
      & \underline{369} [\onlinecite{Smith:1976}] & 
      & 376  & 353  & 367  \\
      5 & $E_\text{fcc} - E_\text{hcp}$ & meV & 0.72 
      & & \underline{14} [\onlinecite{Althoff:PhysRevB.48.13253}]
      & 4  & 4  & 15  & \\
      6 & $E_\text{bcc} - E_\text{hcp}$ & meV & 0.72
      & & \underline{29} [\onlinecite{Althoff:PhysRevB.48.13253}] 
      & 34  & 30  & 18  & \\
      7 & $C_{11}$ & kbar & 1.0 
      & \underline{635} [\onlinecite{Smith:1976}] & 
      & 606  & 602  & 618  & \\
      8 & $C_{12}$ & kbar & 1.0 
      & \underline{260} [\onlinecite{Smith:1976}] &
      & 274  & 237  & 259  & \\
      9 & $C_{13}$ & kbar & 1.0 
      & \underline{217} [\onlinecite{Smith:1976}] &
      & 250  & 219  & 219  & \\
      10 & $C_{33}$ & kbar & 1.0 
      & \underline{665} [\onlinecite{Smith:1976}] &
      & 631  & 623  & 675  & \\
      11 & $C_{44}$ & kbar & 1.0 
      & \underline{184} [\onlinecite{Smith:1976}] &
      & 151  & 155  & 182  & \\
      12 & $\gamma[(0001)]$ & mJ/m$^2$ & 1.0 
      & 680 [\onlinecite{Tyson:1977}]
      & \underline{637} & 583  & 595  & 495  & 310 \\
      13 & $\gamma[(10\bar{1}0)]$ & mJ/m$^2$ & 1.0 
      &     & \underline{721} & 625  & 645  &     &  &  \\
      14 & $E_{\text{sf}}[I_1]$ & eV & 0.1 
      & & \underline{18} & 8  & 7  & 27  & 4 \\
      15 & $E_{\text{sf}}[I_2]$ & eV & 0.1 
      & & \underline{37} & 15  & 15  & 54  & 8 \\
      16 & $E_{\text{sf}}[T_2]$ & eV & 0.1 
      & & \underline{45} & 15  & 15  & & \\
      17 & $E_{\text{sf}}[E]$ & eV & 0.1
      & & \underline{61} & 23  & 22  & & 12 \\
      28 & $E_\text{vac}$ & eV & 1.0 
      & 0.58 $\sim$ 0.89 & \underline{0.82} & 0.58  & 0.56 
      & 0.87 & 0.59 \\
      19 & $\Delta \Avec{F}$ (100~K) & \% & 1.0 & 
      & \underline{0.0} & 38.13 & 201.51 \\
      20 & $\Delta \Avec{F}$ (300~K) & \% & 1.0 & 
      & \underline{0.0} & 29.67 & 93.73 \\
      21 & $\Delta \Avec{F}$ (500~K) & \% & 1.0 &
      & \underline{0.0} & 25.18 & 59.17 \\
      22 & $\Delta \Avec{F}$ (800~K) & \% & 1.0 &
      & \underline{0.0} & 25.08 & 98.77 \\
      23 & $\Delta \Avec{F}$ (1000~K) & \% & 1.0 &
      & \underline{0.0} & 26.93 & 85.31 \\
      24 & $\Delta \Avec{F}$ (1200~K) & \% & 1.0 &
      & \underline{0.0} & 27.30 & 79.77 \\
    \end{tabular}
  \end{ruledtabular}
  \footnotetext[1]{MEAM potential from the present work}
  \footnotetext[2]{MEAM potential from 
    Ref.~\onlinecite{Jelinek:2007:MEAM:Mg-Al}}
  \footnotetext[3]{EAM potential from Ref.~\onlinecite{Liu:1996:EAM:Mg}}
  \footnotetext[4]{Analytic MEAM potential from 
    Ref.~\onlinecite{Hu:2001:AMEAM}}
\end{table*}

Table~\ref{tab:objectives} lists various materials properties of Mg
selected as the objectives to be optimized in constructing Mg MEAM
potential, along with experimental data and \textit{ab initio} data.
It also shows how well each objective has been optimized.  Results
from other previously published Mg potentials are also listed in the
table for comparison.  Table~\ref{tab:objectives} also shows the
weight of individual objectives $w_i$ chosen to optimize the present
potential.  The underlined quantities are the target values chosen for
the MOO procedure.  Whenever possible, the experimental values are
chosen as the target values. If the experimental values are not
available or known to be unreliable, the computed values from the
first-principles method are used.

\begin{figure}[!tbp]
  \includegraphics[width=1\columnwidth]{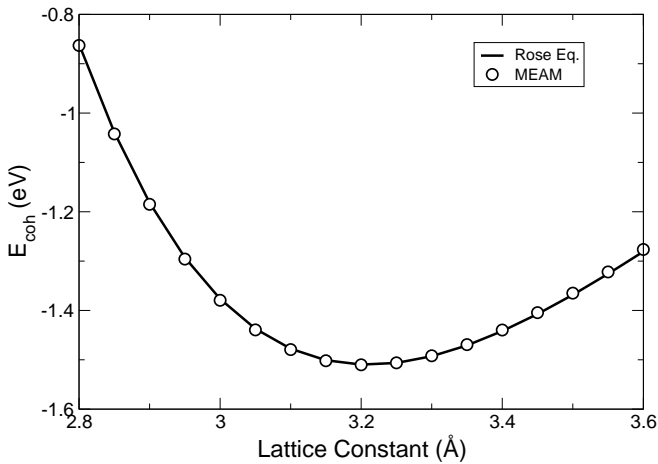}
  \caption{\label{fig:E-V-Mg} The cohesive energies as a function of
    the lattice constant $a$ for Mg atoms in hcp crystal structure
    compared with the ones obtained from the Rose equation.  The data
    points are computed with the present MEAM potential while the
    curve is obtained from the Rose equation.
}
\end{figure}

The present MEAM potential reproduces the experimental lattice
constant, the $c/a$ ratio, and the cohesive energy near perfectly.
Fig.~\ref{fig:E-V-Mg} shows the cohesive energies of Mg atoms in hcp
crystal structure compared with those obtained from the Rose universal
equation of state\cite{Rose:1984:PhysRevB.29.2963} based on the
experimental lattice constant, cohesive energy and bulk modulus.  It
shows a good agreement between the two sets of data.  We also note
that the sequence of the structures is predicted correctly in the
order of stability by the present Mg MEAM potential as shown in
Table~\ref{tab:objectives}.

The surface formation energies of the two common low-index surfaces of
hcp Mg crystals are in good agreement with the experimental values,
representing a significant improvement over the previously published
MEAM potentials.\cite{Jelinek:2007:MEAM:Mg-Al, Lee:2003:MEAM,
  Hu:2003:AMEAM}

As pointed out by Liu et al.\cite{Liu:1996:EAM:Mg}, the stacking fault
energies are difficult quantities for an emprirical potential to
reproduce because they only depend on long range interactions beyond
second nearest-neighbor distances in hcp crystals.  The present MEAM
potential shows a substantial improvement over the previously
published MEAM potential by Hu et al.\cite{Hu:2001:AMEAM} The stacking
fault energies are consistently underestimated by the present MEAM
potential compared to the results of the DFT calculations, while the
results by the EAM potential from Ref.~\onlinecite{Liu:1996:EAM:Mg}
are consistently overestimated.  Table~\ref{tab:objectives} also shows
that the formation energy of single vacancies from DFT calculation is
reproduced quite reasonably by the present MEAM potential.

Table~\ref{tab:objectives} also shows the force-matching against the
\textit{ab initio} forces database.  It shows that the MEAM potential
from the present work reproduces more accurate forces on atoms
compared to the previous MEAM potential\cite{Jelinek:2007:MEAM:Mg-Al}.

\subsection{Additional materials properties}

\begin{table}[!tbp]
  \caption{\label{tab:properties} The additional materials properties 
    of Mg that are not used as objectives for the constructuction of 
    the potential. 
    Comparisons are made to other Mg potentials and experiments.
    $E_{\text{ads}}$ is the adsorption energy; 
    $E_\text{f}^\text{int}$ is the formation energies of interstitial 
    point defects. All energy values are given in eV.
  }
  \begin{ruledtabular}
    \begin{tabular}{lccccccc}
      Property & DFT & MEAM\footnotemark[1]
      & Jelinek\footnotemark[2] \\
      \hline
      $E_{\text{ads}}[(0001)]$ & -0.81 &  -1.46 & -1.50 \\
      $E_{\text{ads}}[(10\bar{1}0)]$ & -1.21 & -1.52 & -1.56 \\
      $E_\text{f}^\text{int}$ (octahedral) & 2.36 & 1.20 & 1.29 \\
      $E_\text{f}^\text{int}$ (tetrahedral) & 2.35 & 1.41 & 1.53 \\
    \end{tabular}
  \end{ruledtabular}
  \footnotetext[1]{MEAM potential from the present work}
  \footnotetext[2]{MEAM potential from 
    Ref.~\onlinecite{Jelinek:2007:MEAM:Mg-Al}}
\end{table}
To validate the present MEAM potential further, we calculated a few
additional materials properties of Mg that were not used as objectives
during the construction of the potential. We obtained the adsorption
energies of a single Mg atom on different surfaces and the formation
energies of interstitial defects as listed in
Table~\ref{tab:properties}.

The adsorption energy of a single adatom $E_{\text{ads}}$ is given by
\begin{equation}
  E_{\text{ads}} = E_{\text{tot}} - E_{\text{surf}} - E_{\text{atom}},
\end{equation}
where $E_{\text{tot}}$ is the total energy of the structure with the
adatom adsorbed on the surface, $E_{\text{surf}}$ is the total energy
of the surface without the adatom, and $E_\text{atom}$ is the total
energy of an isolated atom.  On both $(0001)$ and $(10\bar{1}0)$
surfaces, we placed a single Mg atom at the site where the atoms of
the next layer would normally sit.  The structures were then relaxed
to determine the adsorption energies.  Table~\ref{tab:properties}
shows that the adsorption energies on two Mg surfaces are quite well
reproduced by the present MEAM potential.  The present Mg potential
gives slightly better adsorption energies than the previously
published MEAM potential \cite{Jelinek:2007:MEAM:Mg-Al}.

The formation energy of an interstitial point defect
$E_{\text{f}}^{\text{int}}$ is given by
\begin{equation}
  E_{\text{f}}^{\text{int}} = E_{\text{tot}}[N+1] - (E_{\text{tot}}[N] 
  + \varepsilon),
  \label{eq:Ef_int}
\end{equation}
where $E_\text{tot}[N]$ is the total energy of a system with $N$ Mg
atoms, $E_\text{tot}[N+1]$ is the total energy of a system with $N$
atoms plus one Mg atom inserted at one of the interstitial sites, and
$\varepsilon$ is the total energy per Mg atom in its most stable bulk
structure.  Interstitial atom formation energies were calculated for
Mg at octahedral and tetrahedral sites.  Atomic position and volume
relaxation were performed.  The results of these calculations are
listed in Table~\ref{tab:properties}, to be compared with the results
from the DFT calculations. The present MEAM potential predicts correct
signs for these energies although the magnitudes are about half of
those predicted by DFT. MEAM potentials predict that the octahedral
site will be more stable than the tetrahedral site, while the DFT
calculations indicate that both sites will have nearly the same
formation energies.

\subsection{Thermal properties}

\begin{figure}[!tbp]
  \includegraphics[width=1\columnwidth]{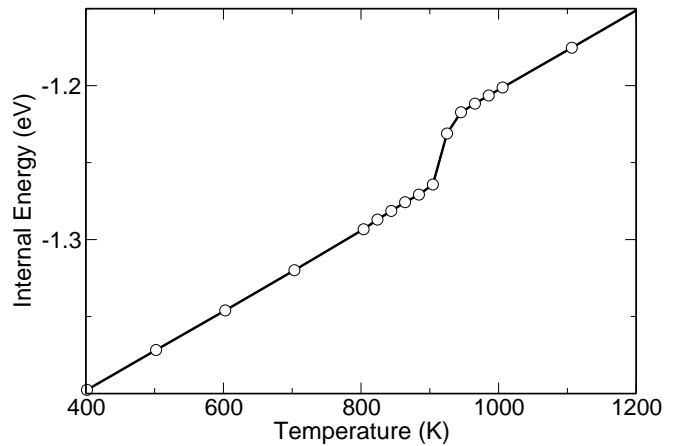}
  \caption{\label{fig:energy-temp} The internal energies of Mg crystal
    in hcp structure as a function of temperature.  The energies are
    obtained from the ensemble average of the MD simulations of five
    structures containing 448 Mg atoms.}
\end{figure}

To validate the new potential for molecular dynamics simulations, we
calculated the melting temperatures of pure Mg crystals. We followed a
single-phase method as described by Kim and
Tom{\'a}nek,\cite{Kim:1994:PRL} in which the temperature is increased
at a constant rate and the internal energy of the system is monitored.
Fig.~\ref{fig:energy-temp} shows the internal energies of Mg crystal
in hcp structure as a function of temperature.  The plot was obtained
from the ensemble average of five hcp structures containing 448 Mg
atoms.  The initial velocity vectors were set randomly according to
the Maxwell-Boltzmann velocity distribution at $T = 100$~K.  The
temperature of the system was controlled by using a Nos\'e-Hoover
thermostat.\cite{Nose:1984, Hoover:1985:PhysRevA.31.1695} It is
clearly seen from Fig.~\ref{fig:energy-temp} that the internal energy
curve makes an abrupt transition from one linear region to another,
marking the melting point.  Using this method, we obtained 920~K as
the melting temperature of Mg crystals.  This result is in good
agreement with the experimental value of 923~K.  Our result represents
a substantial improvement in accuracy from 745~K obtained from a
previously published EAM potential \cite{Liu:1996:EAM:Mg} or 780~K
from a MEAM potential.\cite{Jelinek:2007:MEAM:Mg-Al}

\section{Conclusions}
\label{sec:Conclusion}

In this study, we developed a multi-objective optimization procedure
to construct MEAM potentials with minimal manual fitting.  We
successfully applied this procedure to develop a set of MEAM
parameters for Mg interatomic potential based on first-principles
calculations within DFT.  The validity and transferability of the new
MEAM potentials were tested rigorously by calculating the physical
properties of the Mg systems in many different atomic arrangements
such as bulk, surface, point defect structures, and molecular dynamics
simulations.  The new MEAM potential shows a significant improvement
over previously published potentials, especially for the atomic forces
and melting temperature calculations.

\section{Acknowledgment}

This work has been supported in part by the US Department of Energy
under Grant No.~DE-AC05-00OR22725 subcontract No.~4000054701.
Computer time allocation has been provided by the High Performance
Computing Collaboratory (HPC$^2$) at Mississippi State University.

\bibliography{MEAM,DFT,MD,Mg}

\end{document}